\documentclass[twocolumn,aps,prl,floatsfix,showpacs]{revtex4}
\usepackage{graphics}
\usepackage{amssymb,amsmath}
\usepackage{array}
\usepackage{dcolumn}

\begin{document}

\title{Influence of the incommensurability in
$S\!r_{14-x}C\!a_xC\!u_{24}O_{41}$ family compounds.}

\author{Alain Gell\'e, Marie-Bernadette Lepetit}
\affiliation{Laboratoire de Physique Quantique, IRSAMC~/~UMR~5626,
Universit\'e Paul Sabatier, 118 route de Narbonne, F-31062 Toulouse
Cedex 4, FRANCE}

\date{\today}

\begin{abstract}
The present paper studies the influence of the structural modulation
on the low energy physics of the $S\!r_{14-x}C\!a_xC\!u_{24}O_{41}$
oxides, using ab-initio determination of the on-site and nearest
neighbor effective parameters. The structural modulations appears to
be the key degree of freedom, responsible for the low energy
properties, such as the electron localization, the formation of
dimers in the $x=0$ compound or the anti-ferromagnetic order in the
$x=13.6$ compound.
\end{abstract}


\maketitle

%------------------------------ intro

The family of $S\!r_{14-x}C\!a_xC\!u_{24}O_{41}$ transition-metal
oxides has attracted a lot of attention in the last decade. Indeed,
the superconducting state observed in the $x>11.5$ compounds under high
pressure~\cite{ObsSupra}, is supposed to be the realization of the
remarkable theoretical prediction of superconductivity in two-legs 
doped spin ladders~\cite{TheoEchSupra}.
Moreover, this family of compounds exhibits a large diversity in
electric and/or magnetic properties when chemical ---~isovalent
substitution of $S\!r$ by $C\!a$~--- and physical pressure are
applied. For instance, under applied pressure, the compounds change
from semi-conductor to conductor and finally to superconductor.

This family of compounds possesses a complex layered structure of two
alternating subsystems~\cite{struc1}. The layers of the first subsystem are
composed of weakly-coupled $C\!uO_2$, spin-$1/2$ chains along the
${\bf c}$ direction. The spins, supported by $3d$ orbitals of the
$C\!u^{2+}$ ions, are coupled via two $90^\circ$ $C\!u$--$O$--$C\!u$
bonds. The layers of the second subsystem are composed of
weakly-coupled, two-leg spin-1/2 ladders also along the ${\bf  c}$
direction. The spins are strongly antiferromagnetically coupled on
both legs and rungs, due to the $180^\circ$ $C\!u$--$O$--$C\!u$ bonds.
The cell parameters of the two subsystems, in the direction of both
chains and ladders, are incommensurate. The compounds have a
pseudo-periodicity of 10 chain units for 7 ladder units.

Electro-neutrality analysis shows that these systems are intrinsically
doped with six holes by formula unit (f.u.). Similar to high-$T_c$
superconductors, the holes are expected to be mainly supported by the
oxygen $2p$ orbitals and to form Zhang-Rice~\cite{ZR} singlets with
the associated-copper hole. NEXAFS experiments~\cite{XRay00} have
later supported this assumption. A calculation of the Madelung
potential~\cite{Mad97} on the concerned oxygen sites suggests that for
the undoped compound, $S\!r_{14}C\!u_{24}O_{41}$, the chains exhibit
a larger electro-negativity than the ladders, resulting in a
localization of all the holes on the former. Different
experiments~\cite{COpt97,XRay00}, however suggest that about one hole
per f.u. is located on the ladders. Under $C\!a$ substitution, the
same experiments, as well as $C\!u$ NMR~\cite{RMN98}, show a transfer
of part of the holes to the ladders. However the precise number of
transferred holes is still under debate.  X-ray data~\cite{XRay00}
suggest a small hole transfer to the ladders ($1.1$ for $x=12$), while
optical conductivity~\cite{COpt97} and $^{63}C\!u$ NMR
studies~\cite{RMN98} show a larger hole transfer, of respectively
$2.8$ ($x=11$) and $3.5$ ($x=11.5$).

Let us first focus on the undoped system, $S\!r_{14}C\!u_{24}O_{41}$.
This is a semiconductor with a $0.18\ eV$ gap.
The spin ladders have a
singlet ground state with a spin gap of about $35-47\
meV$~\cite{Neut96,RMN98,RMN97}.
Surprisingly the spin chains also
exhibit a singlet ground state with a spin gap of $11-12\
meV$~\cite{Magn96B,RMN97,RMN98B,Neut98,Thermo00}. Since homogeneous
spin chains are known to be gap-less in the spin channel, the
existence of a gap witnesses their strongly inhomogeneous character.
In fact, the electronic structure of the $S\!r_{14}C\!u_{24}O_{41}$
chains is usually understood as formed by weakly interacting
dimmers~\cite{Magn96,Magn96B,ESR96,Neut96}. It is now well established
that these dimers are formed by second-neighbor spins separated by a
Zhang-Rice singlet (ZRS), and order~\cite{RMN98B,ESR01,Neut98,Neut99}
according to a pseudo periodicity of 5 sites (one dimer followed by
two ZRS).

When $S\!r$ is substituted by $C\!a$ the system becomes more
metallic. The doped compound with $x=10$ shows a gap of only $0.023
eV$ ~\cite{Magn96B}. This increased conductivity, as a function of the
$C\!a$ doping, is usually understood as a consequence of the hole
transfer from the chains to the ladders, in which the conduction is
supposed to occur. However, a possible enhancement of the holes
mobility within the chains is also evoked~\cite{XRay00,ESR01}. Neutron
scattering experiments~\cite{Neut99} showed that the dimerization
becomes unstable with $C\!a$ substitution and disappears for $x>8$,
although the magnetic interactions within and between the dimers
remain unchanged. In parallel, both ESR~\cite{ESR01} and thermal
expansion data~\cite{Thermo00} witness a progressive disappearance of
the charge order with increasing doping. Finally, at large doping ($x
\ge 11$) and very low temperatures ($<2.5K$) an anti-ferromagnetic
phase is observed~\cite{Magn99,ESR01}.

An important aspect of these compounds, which is most of the time
neglected, is the modulation of the two subsystems. Indeed, the mutual
influence of the two subsystems results in a modulation of each of
them with the periodicity of the other. These distortions are
particularly large on the chains. Indeed, in the highly doped systems,
the $C\!u$--$O$--$C\!u$ angle varies between $89^\circ$ and
$99^\circ$, while the $C\!u$--$O$ distance varies with an amplitude of
$19\%$. The magnetic interactions being very sensitive to both bond
angles and distances between the magnetic sites and the bridging
ligands, one can expect that the modulations will be of importance for
the low energy physics of the compounds. In the ladder subsystem, the
structural distortions are of weaker amplitude since the
alkaline-earth counter-ions are attached to it. In addition, the
$C\!u$--$O$--$C\!u$ angle varies around $\theta = 180^\circ$ and therefore 
super-exchange mechanism (that scales as $\cos^4{\theta}\simeq
1-2\theta^2$) should be dominant. Thus, while the effect of the structural
modulations on the ladders may be of importance, it can be expected to
be much weaker than on the chains sub-system.

The aim of this paper is to study to which extent the modulations
influence the chains electronic structure. For this purpose, we
performed ab-initio calculations so that to accurately evaluate the
influence of the modulations on the magnetic orbital energies (OE) and
nearest neighbors (NN) interactions. We choose for these calculations
the $S\!r_{14}C\!u_{24}O_{41}$ and
$S\!r_{0.4}C\!a_{13.6}C\!u_{24}O_{41}$ compounds in their low
temperature phase~\cite{struc}.

%------------------------------ method
As the interactions between magnetic sites are essentially local, they
can be accurately determined using an embedded fragment ab-initio
spectroscopy method~\cite{revue}. The fragment includes the magnetic
centers, the bridging oxygens mediating the interactions, and their
first coordination shell. Short-range crystal effects are thus treated
explicitly while the long-range crystal effects, such as the Madelung
potential, are treated within an appropriate bath.  The calculations
have been performed using the DDCI method~\cite{DDCI}, a selected
multi-reference single and double configuration interaction that
properly treats (i) the strongly correlated character of the system,
(ii) the mediation of the interaction via the bridging ligands and
(iii) the screening effects on these processes~\cite{bridge}. The
reference space has been chosen to be composed by the copper magnetic
orbitals. The basis sets are of valence $3\zeta$~\cite{bases} quality
for $C\!u$ and $2\zeta + p$ for $O$.
This method has proved to be very accurate in the determination of the
the local effective interactions as well as local electronic
structures. One can cite, for instance, the remarkable results
obtained on the effective exchange and hopping determination in
high-$T_c$ parent compounds~\cite{DDCIhtc}, copper
chains~\cite{DDCIcha} and ladder~\cite{DDCIlad} systems, where the
computed values are within experimental accuracy, as
well as on the charge ordering in the sodium vanadate low temperature
phase~\cite{vana}.

Singlet-triplet excitation energies on embedded $C\!u_2 O_6$ fragments
will thus yield the effective exchange integrals, while the first
doublet-doublet excitation energies and associated wave functions
yield both magnetic (hole) OE and hopping effective integrals. 
In order to study the influence of the incommensurability on these
parameters, we performed calculations on 11 fragments associated
to successive cells of the chain sub-system. 

%------------------------------ result

%%%% Energies sur site
\begin{figure}[h] %%%%% variation E
\resizebox{8cm}{3cm}{\includegraphics{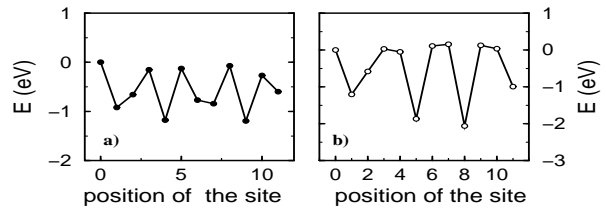}}
%\vspace*{0.5eM}
\caption{Energy of the hole orbital as a function of its 
position along the chain. a) for $x=0$, b) for
$x=13.6$.}
\label{E1}
\end{figure}
Let us first look at the hole OE (fig.~\ref{E1}). One
sees immediately that their modulations  are very
large. Indeed, in the non-doped system, the OE vary
within a range of $1.2\ eV$. In the $C\!a$ doped system, the variation
is even much larger and spans a $2.2\ eV$ range. Let us notice, that
in both systems, the OE variation is larger than the
hopping and exchange energy scales (respectively of $150\ meV$ and
$20\ meV$, see below). Thus, this is the OE that will
dominate the low energy physics through the localization of the
magnetic electrons (holes) on the low (high) energy sites.
One should notice that the variations of the OE are due
to the crystal distortions. Indeed, when using the average crystal
structure ---~where the subsystems distortions have been omitted~---
the OE exhibit only very small variations~\cite{env}
($<30\ meV$ for the undoped compound).

Before studying the electrons (holes) localization along the chain, let
us take a look at the variations of the NN hopping and exchange
interactions.

%%% Echange 
\begin{figure}[h] \vspace*{2ex}
\resizebox{8cm}{5cm}{\includegraphics{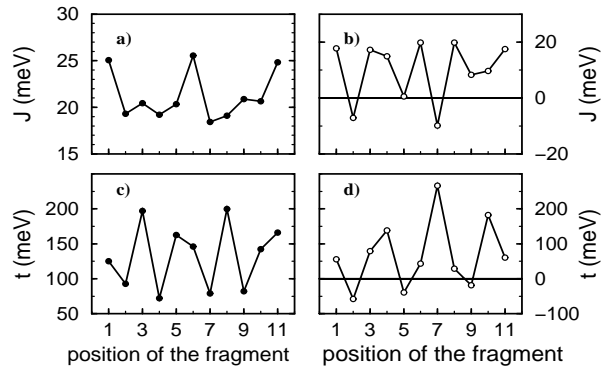}}
%\vspace*{-1eM}
\caption{Exchange $J$ and hopping $t$ integrals as a function of the
position of the fragment along the chain. a) $J$ for $x=0$, b) $J$ for
$x=13.6$, c) $t$ for $x=0$, d) $t$ for $x=13.6$.}
\label{J1}
\end{figure}
The effective exchange integrals, obtained for the 11 fragments,
are reported in fig.~\ref{J1}. As expected the NN effective
exchange for the $x=0$ compound is ferromagnetic and exhibits small
modulations around an average value of $21.3\ meV$, the standard
deviation being $2.5\ meV$. On the contrary, the $x=13.6$ compound
does not follow the expectations. Indeed, the effective exchange
varies greatly, going from ferromagnetic values ($20\ meV$) up to
anti-ferromagnetic interactions as large as $-10\ meV$. These
anti-ferromagnetic interactions are observed either for large
$C\!u$--$O$--$C\!u$ angles ($\simeq 98^\circ$) or for large angles
between the two magnetic orbitals ($>10^\circ$) associated with
short $C\!u$--$C\!u$ distances. In both cases, these strong
distortions allow super-exchange mechanism to take place.

%%% saut 
The effective hopping integrals between NN copper atoms are expected
to be quite small, since the nearly $90^\circ$ $C\!u$--$O$--$C\!u$
angles forbid the through-bridge contribution via the oxygen orbitals.
However, this is not what is observed in our calculations (see
fig.~\ref{J1}) since the hopping integrals present very large
modulations and can reach values as large as $208\ meV$ for $x=0$ and
$266\ meV$ for $x=13.6$. These amplitudes are as large as $1/3$ of the
hopping observed in systems with $180^\circ$ $C\!u$--$O$--$C\!u$
angles, such as high-$T_c$ superconductors~\cite{DDCIhtc} or $C\!uO_3$
corner-sharing chains~\cite{DDCIcha}.

%%%% Extrapolations et ordre de charge 
At this point it is clear that, the systems modulations, and the
variations of these modulations according to the $C\!a$ doping, are
crucial for the low energy properties of this family of iso-electronic
compounds. We will now study whether they can explain the existence of
the dimerization that is observed in the weakly doped compounds but
not in the highly doped ones. For this purpose we need to extrapolate
the values of the OE over the whole chain. It is usual to do so using
a Bond Valence Sum (BVS) analysis, that analyzes the distances between
the metal atom and its first coordination shell. However, for the
$S\!r_{0.4}C\!a_{13.6}C\!u_{24}O_{41}$ system, the BVS analysis yields
quite different results from the OE. A further analysis discloses
that, in these systems, the Madelung potential on the magnetic
centers, and thus the OE, is sensitive to the atomic displacements not
only of the first coordination shell, but up to the $8^{\rm th}$ shell
of neighbors. This is for this reason that the BVS analysis fails to
correctly reproduce the copper valence for this type of compounds.

\begin{figure}[h] %%%%% Analyse de Fourier des energies orbitalaires
\resizebox{8.cm}{3cm}{\includegraphics{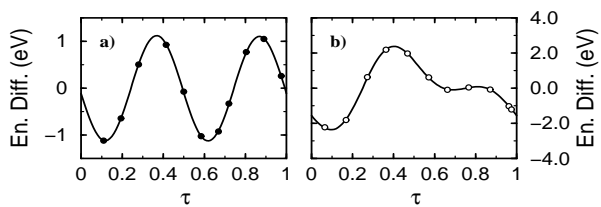}}
\caption{ Orbital-energy differences as a function of the fourth
coordinate $\tau$, a) for $x=0$, b) for $x=13.6$. Solid line represent
the fit.}
%Fourier analysis of the computed orbital-energy differences
%between neighboring sites, a) for $x=0$, b) for $x=13.6$.. Circles
%represent the computed values, and the solid line the fit.}
\label{four}
\end{figure}

In order to extrapolate the OE over the whole chain, we used the
crystallographic description of the incommensurate structure in a four
dimensional space. Each subsystem is thus described by the three $a$,
$b$ and $c$ usual spatial dimensions and a fourth coordinate $\tau$
that has the periodicity of the other subsystem and describes the
modulations~\cite{struc4}.  We fitted the computed values using a
Fourier analysis as a function of $\tau$ (see fig.~\ref{four}).  One
can note that the doped and undoped compounds present very different
OE curves as a function of $\tau$. Let us notice on
fig.~\ref{four}a the half periodicity of the magnetic cell for the
$S\!r_{14}C\!u_{24}O_{41}$ compound, observed in neutrons scattering
experiments~\cite{Neut98,Neut99}.

\begin{figure*} %% remplissages
\resizebox{17.9cm}{4.4cm}{\includegraphics{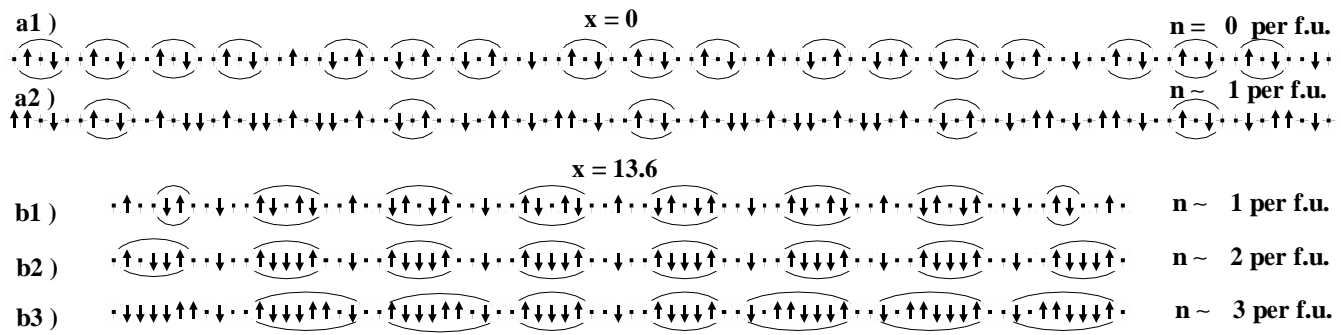}}
%\vspace*{0.5eM}
\caption{Localization of the magnetic electrons along the chain, as a
function of the number $n$ of holes transfered to the ladders. a) for
$x=0$, b) for $x=13.6$. The number of sites considered in both cases
corresponds to a better approximation of the compounds unit cells than
the 10/7 ratio, that is $c_{ch}/c_{la} =3.9235/2.7268 \simeq 100/69.5$
for $x=0$, and $c_{ch}/c_{la} =3.90136/2.73607 \simeq 77/54$ for
$x=13.6$. Dots stand for ZR singlets, and the ellipsoids delimit
spin clusters, which are magnetically very stable (with at most one
second neighbor frustrated interaction).}
\label{rempl}
\end{figure*}
Figure~\ref{rempl} reports the electrons (holes) localization along the
chain as derived from the extrapolated on-site orbital energies.
In the $x=13.6$ case, a Fourier analysis has also been done on the
effective NN exchange in order to predict the sign of the NN
interactions for the different positions along the chain. The second
neighbor exchange has been considered to be anti-ferromagnetic in
agreement with experimental results.

For the undoped $S\!r_{14}C\!u_{24}O_{41}$ system we studied the two
types of filling considered in the literature, namely with all the
holes on the chains (fig.~\ref{rempl}a1) and with one hole transfer
per f.u. (fig.~\ref{rempl}a2). When all the holes are on the chains,
we retrieve a spin arrangement consistent with the experimental
observations~\cite{Neut98,Neut99,RMN98}, that is dimeric units formed
of two second neighbors spins, separated by two ZR singlets. These
dimers are clustered by three or four units separated by a free spin,
that is a spin with neither first nor second neighbor spin. We found 5
free spins for 100 sites ($0.5$ per f.u.) to be compared with the
magnetic susceptibility measurements of $0.55$ free spins per
f.u.~\cite{Magn96B}. The number of dimers is 17 ($1.7$ per f.u.) to be
compared with the magnetic susceptibility finding of $1.47$ per
f.u.. When one hole per f.u. is transfered to the ladders, the picture
is totally modified. Indeed, the free spins totally vanish and first
neighbor ferro-magnetically coupled spins appear. At the same time the
number of (isolated) dimers is strongly reduced, with only 5 for 100
sites, that is $0.5$ per f.u.  From this analysis it is clear that if
there are holes transferred to the ladders (at low temperature) in the
undoped compound, this number is much smaller than 1 per f.u..

For the $x=13.6$ doped system we studied three types of filling where
$n=$ 1,2, or 3 holes have been transferred to the ladders. For all
fillings, we retrieve arrangements with nearest neighbors spins and no
second neighbors dimeric units. The chains are essentially composed of
low-spin clusters (either in a singlet or doublet state), with spin
arrangements that present  weak exchange frustration. Such
fillings are thus expected to be specially stable, and should be put
into perspective with the anti-ferromagnetic ordering seen in magnetic
susceptibility and ESR measurements. For $n=1,2$ one still observes a
large number of free spins, while for $n=3$ they
have essentially disappeared, in agreement with magnetic
susceptibility experiments~\cite{Magn99}.

%%%%% conlusion 
 
In summary, we have studied the importance of the structural
modulations on the low energy physics of the
$S\!r_{14-x}C\!a_{x}C\!u_{24}O_{41}$ family. Surprisingly these
distortions are not simply responsible for parameters modulations
around their average value (except for the NN effective exchange on
the undoped compound), but induce very large variations of the orbital
energies, NN effective hopping and exchange. This is the variation of
the OE (which spans a range of $1.2\ eV$ for $x=0$ and
$2.2\ eV$ for $x=13.6$) that is responsible for the low energy
properties of the compounds, through the localization of the magnetic
electrons. It is in particular responsible of the formation of dimers
in the undoped compound. In view of these results one can also suppose
that the stabilization of part of the chains sites by larger
structural modulations, is responsible for the hole transfer toward
the ladders in the doped compounds. In the $x=13.6$ compound, the
structural modulation is so large that it even reverse the sign of the
NN effective exchange on part of bonds, lifting the exchange
frustration that would arise for the large chain filling. In
conclusion, one can say that the structural modulation is the key
parameter, responsible for the large variation of the low energy
properties in this family of compounds.

{\bf Acknowledgment :} the authors thank Dr. D. Maynau for providing
us with the CASDI suite of programs.

 \end{document}